\documentclass[reprint,amsmath,amssymb,aps,prb,showpacs,floatfix]{revtex4-1}

\usepackage{graphicx} 
\usepackage{bm}
\usepackage{subfigure}
\usepackage{multirow}
\usepackage{verbatim}
\usepackage{longtable}

\newcommand{\e}{e}
\renewcommand{\i}{i}
\renewcommand{\figurename}{Fig.}
\renewcommand{\tablename}{Table}
\newcommand{\sectionname}{Section}

\newcommand\abs[1]{\lvert#1\rvert}
\newcommand\absB[1]{\left\lvert#1\right\rvert}

\newcommand\bra[1]{\langle#1\rvert}
\newcommand\ket[1]{\lvert#1\rangle}
\newcommand\braket[2]{\langle#1\rvert#2\rangle}

\newcommand{\Arg}{\mathrm{Arg}}

\newcommand{\ve}{\varepsilon}
\newcommand{\vphi}{\varphi}

\newcommand{\m}{\mathcal}

\newcommand{\up}{\uparrow}
\newcommand{\down}{\downarrow}

\newcommand{\bk}{\mathbf{k}}
\newcommand{\br}{\mathbf{r}}
\newcommand{\bp}{\mathbf{p}}

\newcommand{\bB}{\mathbf{B}}

\newcommand{\cd}{c^{\dag}}
\newcommand{\can}{c^{\phantom{\dag}}}
\newcommand{\dd}{d^{\dag}}
\newcommand{\dan}{d^{\phantom{\dag}}}

\newcommand{\SO}{\mathrm{SO}}
\newcommand{\orb}{\mathrm{orb}}
\newcommand{\LR}{\mathrm{LR}}
\newcommand{\LRo}{\mathrm{LR}}
\newcommand{\D}{\mathrm{D}}
\newcommand{\T}{\mathrm{T}}
\newcommand{\mrL}{\mathrm{L}}
\newcommand{\mrR}{\mathrm{R}}
\newcommand{\s}{s}

\newcommand\sgn{\mathrm{sgn}}

\begin{document}

\title{Cotunneling renormalization in carbon nanotube quantum dots}

\author{Gediminas Kir\v{s}anskas}
\affiliation{Center for Quantum Devices, Niels Bohr Institute, University of Copenhagen, DK-2100 Copenhagen \O, Denmark}

\author{Jens Paaske}
\affiliation{Center for Quantum Devices, Niels Bohr Institute, University of Copenhagen, DK-2100 Copenhagen \O, Denmark}

\author{Karsten Flensberg}
\affiliation{Center for Quantum Devices, Niels Bohr Institute, University of Copenhagen, DK-2100 Copenhagen \O, Denmark}

\date{\today}

\begin{abstract}
We determine the level shifts induced by cotunneling in a Coulomb blockaded carbon nanotube quantum dot using leading-order quasi-degenerate perturbation theory within a single nanotube ``shell.'' It is demonstrated that otherwise degenerate and equally tunnel coupled $K$ and $K'$ states are mixed by cotunneling and therefore split up in energy except at the particle-hole symmetric midpoints of the Coulomb diamonds. In the presence of an external magnetic field, we show that cotunneling induces a gate dependent $g$-factor renormalization, and we outline different scenarios which might be observed experimentally, depending on the values of both intrinsic $KK'$-mixing and spin-orbit coupling.
\end{abstract}

\maketitle 

\section{\label{sec:intro}Introduction}

Inelastic cotunneling~\cite{fran2001} provides a powerful spectroscopic probe of the electronic and magnetic states of a Coulomb blockaded nanostructure, be it a quantum dot~\cite{Zum04,Paaske06,jesp2011,Katsaros11} or a single molecule.~\cite{Yu04,Osorio07b,Roch08a,Osorio10,Roch11} The basic cotunneling process itself corresponds to a virtual charge fluctuation of the dot, and as such it also renormalizes the very spectrum which is being investigated.~\cite{wolf1969,hald1978} For sufficiently strong tunnel couplings this may even instigate sharp many-body resonances in the system, such as the celebrated Kondo effect for spinful nano structures,\cite{Goldhaber98,Nygard00} which in itself is a telling spectroscopic signature.

For a dot in a specific charge state, the thresholds for inelastic cotunneling correspond to energy differences between excited states and the ground state. Therefore, a threshold is only renormalized by cotunneling if the energy of the corresponding excited state is renormalized differently from the ground state, i.e., if the two have different tunnel couplings. Furthermore, since fluctuations to the two adjacent charge states are generally different, this often leads to a marked gate voltage dependence of the cotunneling thresholds. This effect was first predicted~\cite{mart2003,sind2007} for systems with ferromagnetic leads, where charge fluctuations become spin dependent and consequently induce a gate dependent exchange field on the dot. Experiments have later confirmed the presence of this, potentially sizeable, exchange field~\cite{pasu2004} as well as its characteristic gate dependence,~\cite{haup2008} and the same effect has recently been shown to have a strong influence also on the measured tunneling magnetoresistance.~\cite{koll2012}

\begin{figure}[!h]
\begin{center}
\includegraphics[width=0.8\columnwidth]{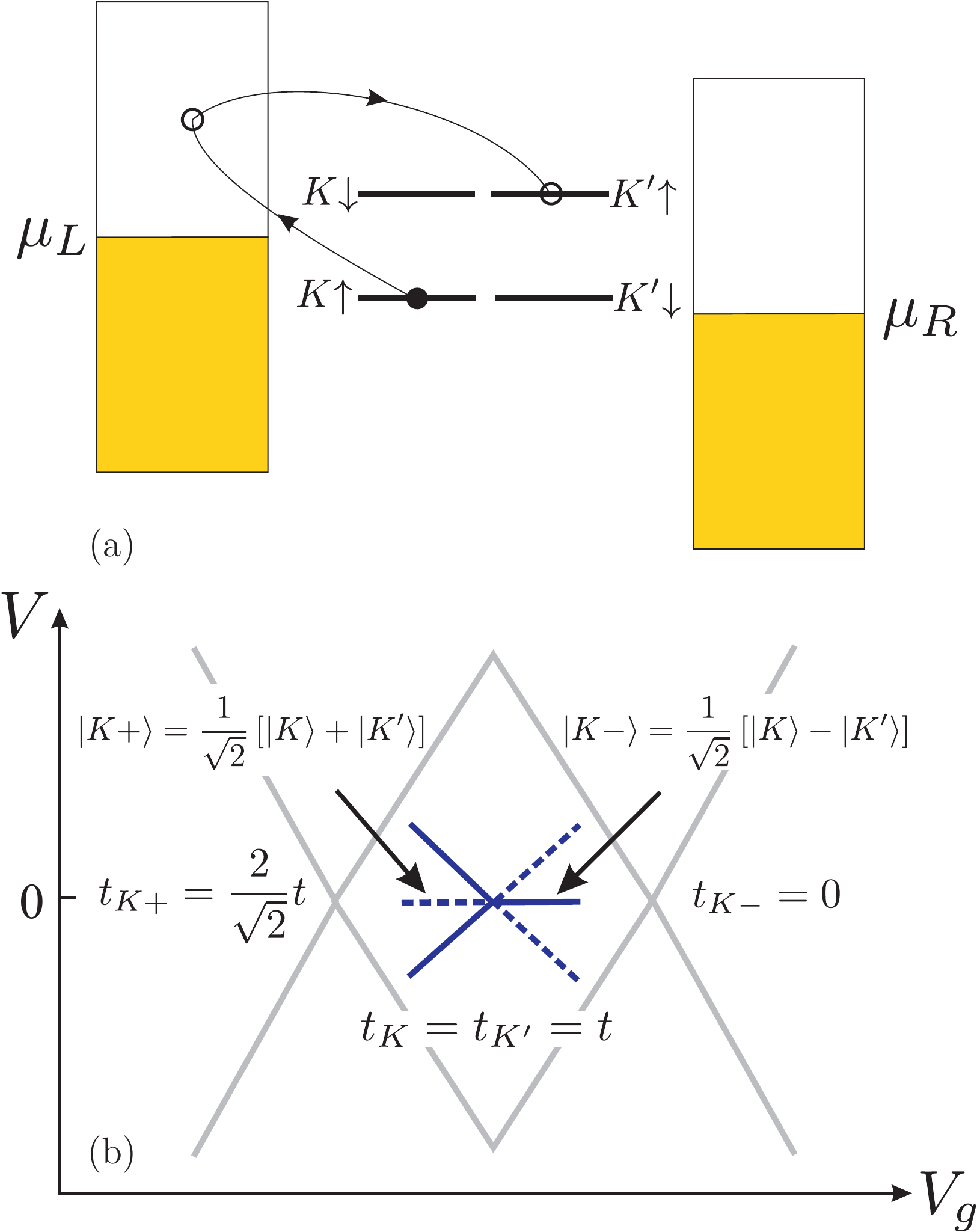}
\caption{\label{fig0} (Color online) Qualitative picture of tunneling induced $KK'$-mixing for a carbon nanotube coupled to normal leads in a zero magnetic $B=0$ field. a) Example of the tunneling process, which introduces $KK'$-mixing with a single electron in a carbon nanotube ``shell''. Here $\mu_{L,R}$ denotes chemical potentials of the leads. b) Schematic of the cotunneling threshold (solid and dashed blue lines) for a single charge Coulomb diamond, in the absence of spin-orbit coupling and disorder induced $KK'$-mixing, and with real tunneling amplitudes $t_{K}$, $t_{K'}$. The effective tunneling renormalized ground state of the quantum dot is $\ket{K+}$ with larger tunneling rate $\Gamma_{K+}\sim\abs{t_{K+}}^2$ (dashed blue line) at the left side of the diamond and $\ket{K-}$ with smaller tunneling rate $\Gamma_{K-}\sim\abs{t_{K-}}^2$ (solid blue line) at the right side.}
\end{center}
\end{figure}

Experiments with carbon nanotube quantum dots contacted by either normal~\cite{holm2008} or superconducting~\cite{grove2009} leads showed a characteristic gate dependence of inelastic cotunneling thresholds repeated over many different charge states. This systematic behavior could again be ascribed to cotunneling renormalization, by invoking different tunneling amplitudes for the two different states involved in a carbon nanotube ``shell.'' Since, however, $K$ and $K'$ states (see below) in a carbon nanotube are time-reversal partners, they will have the same Hamiltonian overlap with the lead states and hence equal tunneling amplitudes. The two differently coupled states observed in Ref.~\onlinecite{holm2008} therefore must correspond to a mixture of $K$ and $K'$, reflected also in the non-zero cotunneling threshold. This in turn implies that the observed gate dependence of thresholds should disappear when applying a strong magnetic field along the tube, again making $K$ and $K'$ good quantum numbers, which indeed was confirmed in a more recent experiment.~\cite{grov2012}

In the present paper, we revisit the problem of cotunneling renormalization in a carbon nanotube quantum dot, and include now the cotunneling induced mixing of the dot states. We also include the effects of spin-orbit coupling as well as intrinsic $KK'$-mixing, $\Delta_{KK'}$, due to impurities and/or nearby gates or substrate breaking the lattice symmetry of the tube. We model the leads by a simple flat band metal which carries no information about the quantum numbers, $K$ and $K'$, of the dot thus giving rise to tunneling induced $KK'$-mixing (see \figurename~\ref{fig0}a). This assumption excludes from our survey the possibility of orbital quasi-degeneracy and the associated $SU(4)$ Kondo effect, which has been discussed in connection to experiments on certain carbon nanotube quantum dots.~\cite{Makarovski06,Makarovski07,Herrero05}

Using quasi-degenerate perturbation theory, we determine the leading-order tunneling renormalization of thresholds for inelastic cotunneling, strictly valid only deep inside the Coulomb diamond delimiting the regions of definite dot charge. Throughout the paper, we neglect the sub-leading effects of level broadening. While formally correct, it should be kept in mind that the tunneling induced fine structure which we report here will eventually be slightly smeared in any experimental realization. A very recent work~\cite{splet2012} has reported interesting effects of tunneling renormalization of the lead-dot tunneling amplitudes themselves, and also this effect will be omitted here as a subleading effect. Figure~\ref{fig0}b illustrates one of our central findings, namely, the tendency for otherwise degenerate dot states to split up away from the particle-hole symmetric midpoint of the Coulomb diamond. Since $K$ and $K'$ states have equal tunnel couplings, the system takes advantage of the tunneling induced $KK'$-mixing allowing to form respectively a strongly coupled ground state, $|K+\rangle$,  near the boundary for removing an electron (left side), and a weakly coupled ground state, $|K-\rangle$, when close to adding an electron (right side). Altogether, this gives rise to a characteristic gate voltage dependent splitting of these nominally degenerate tube states. With outset in this scenario, we will then include the extra effects of spin-orbit coupling, intrinsic $KK'$-mixing, and an applied magnetic field.


The paper is organized as follows. In \sectionname~\ref{sec:model} a constant interaction model for a carbon nanotube quantum dot is introduced. In \sectionname~\ref{sec:s3} the tunneling induced shifts of eigenenergies are considered perturbatively to second order, and in \sectionname~\ref{sec:s4} effects on cotunneling thresholds due to these energy shifts are discussed. In \sectionname~\ref{sec:s5} we examine the magnetic field dependence of the renormalized cotunneling thresholds, and concluding remarks are made in \sectionname~\ref{sec:concl}. A more elaborate discussion of the single particle states and the tunneling amplitudes in a carbon nanotube quantum dot is relegated to \appendixname~\ref{sec:appendixA} and comparison of second- and fourth-order perturbation theory in the tunneling Hamiltonian is provided in \appendixname~\ref{sec:appendixB}. In \appendixname~\ref{sec:appendixR} detailed derivation of Eq. (\ref{soehm}) is given. Throughout the paper, except in \appendixname~\ref{sec:appendixA}, we employ units in which the reduced Planck constant, the elementary charge, and the Bohr magneton are equal to one, i.e., $\hbar=e=\mu_B=1$.

\section{\label{sec:model}Model}

The system under consideration is a quantum dot connected to source and drain electrodes and capacitively coupled to a gate electrode. Such a system can be modeled by the following Hamiltonian
\begin{equation}\label{ham}
H=H_{\LR}+H_{\D}+H_{\T},
\end{equation}
where
\begin{equation}\label{hamLR}
H_{\LR}=\sum_{\alpha \nu\s}\left(\ve_{\alpha\nu}-\mu_{\alpha}\right)\cd_{\alpha\nu\s}\can_{\alpha\nu\s},\\
\end{equation}
describes the source and drain electrodes as two reservoirs of noninteracting electrons. Here $\alpha=L,R$ stands for the left or right lead, $\mu_{L,R}=\pm V/2$ is the chemical potential of the leads, which depends on the applied bias voltage $V$, $\s=\up,\down$ denotes the spin of the electron, and $\nu$ are other quantum numbers of electrons. The quantum dot region is described by $H_{\D}$ and we model it as four localized single particle states with the interaction between electrons assumed to be constant:
\begin{equation}\label{hamD}
H_{\D}=\sum_{n=1}^{4}\ve_{n}\dd_{n}\dan_{n}+U(N-N_g)^2.
\end{equation}
Here $N_g$ corresponds to the gate voltage ($V_g=2UN_g$), $N=\sum_{n=1}^{4}\dd_{n}\dan_{n}$ is the total number of electrons on the dot and $2U$ denotes the total capacitive charging energy of the dot. The single particle spectrum $\ve_n$ for a carbon nanotube quantum dot, which we will be interested in, is defined in \sectionname~\ref{sec:modelA}.

The coupling between the leads and the dot is described by the following tunneling Hamiltonian
\begin{equation}\label{hamT}
H_{\T}=
\sum_{\substack{\alpha\nu\s\\n}}
\left(t_{\alpha\nu \s}^{n}\cd_{\alpha\nu\s}\dan_{n}
+(t_{\alpha\nu \s}^{n})^{*}\dd_{n}\can_{\alpha\nu\s}\right),
\end{equation}
where $t_{\alpha\nu \s}^{n}$ is the tunneling amplitude from the dot state $n$ to the lead state $\alpha\nu\s$. These tunneling amplitudes are defined in \sectionname~\ref{sec:modelB}. We will treat the tunneling Hamiltonian $H_{\T}$ as a perturbation to $H_{\LR}+H_{\D}$, when examining its influence on the spectrum of the dot.

\subsection{\label{sec:modelA}Single particle states and spectrum }

Carbon nanotubes have a nearly fourfold degenerate energy level structure,\cite{cobd2002,liang2002,izum2009,jesp2011} which is due to intrinsic spin ($\up$, $\down$), and so-called isospin or valley index ($K$, $K'$). In a lot of experimental cases every four states,  ``a shell'', are separated by large energy $\Delta{E}$ comparable or bigger than the charging energy $2U$.\cite{jesp2011} The fourfold degeneracy is broken by spin-orbit coupling and disorder. However, if there is no magnetic field, the system still has time-reversal symmetry and we are left with a pair of twofold degenerate states, called Kramers doublets. We label these nearly fourfold degenerate states as
\begin{equation}
\ket{K\up}, \ket{K'\down}, \ket{K\down}, \ket{K'\up},
\end{equation}
which we will refer to as the $KK'$-basis. Without magnetic field the time-reversal partners (Kramers doublets) in the above set are the states
\begin{subequations}
\begin{align}
&\ket{K\up}=T\ket{K'\down}, \\
&\ket{K\down}=T\ket{K'\up},
\end{align}
\end{subequations}
where $T$ is the time-reversal operator.

Following Ref.~\onlinecite{jesp2011}, we will use the following single particle Hamiltonian written in $KK'$-basis for a single ``shell'' of a carbon nanotube quantum dot:
\begin{equation}\label{kkphm}
\begin{aligned}
H_s&=\bordermatrix{&\ket{K\up}   &\ket{K'\down} &\ket{K\down} &\ket{K'\up} \cr
                &E_{+1,+1}    &0             &0            &\frac{1}{2}\Delta_{KK'}\cr
                &0            &E_{-1,-1}     &\frac{1}{2}\Delta^{*}_{KK'} &0   \cr
                &0            &\frac{1}{2}\Delta_{KK'}  &E_{+1,-1}    &0   \cr
                &\frac{1}{2}\Delta^{*}_{KK'} &0             &0            &E_{-1,+1}   }\\
&+\frac{1}{2}g_s B
\bordermatrix{&\ket{K\up}   &\ket{K'\down} &\ket{K\down} &\ket{K'\up} \cr
                &\cos\zeta  &0             &\sin\zeta    &0           \cr
                &0          &-\cos\zeta    &0            &\sin\zeta   \cr
                &\sin\zeta  &0             &-\cos\zeta   &0           \cr
                &0          &\sin\zeta     &0            &\cos\zeta   },
\end{aligned}
\end{equation}
where $\Delta_{KK'}$ is $KK'$-mixing due to disorder, $g_s\approx 2$ is the electron's Land\'{e} $g$-factor, $\zeta$ is the angle between magnetic $\bB$ field and the tube axis, and energies in the diagonal are given by
\begin{equation}\label{esawa}
E_{\tau,s}=\tau s\frac{\Delta_{\SO}}{2}\mp\tau g_{\orb}\frac{B\cos\zeta}{2},
\end{equation}
where $\tau=+1(-1)$ for $K(K')$, $s=+1(-1)$ for spin up (down), $g_{\orb}$ is the effective orbital Land\'{e} $g$-factor, and the upper (lower) sign corresponds to the conduction (valence) band. After diagonalizing the Hamiltonian (\ref{kkphm}), we find the eigenspectrum $\ve_n$. In our discussion of tunneling induced energy shifts, we will use the energy spectrum for the conduction band and assume $\Delta_{\SO}>0$.

\subsection{\label{sec:modelB}Tunneling amplitudes}

To define the tunneling amplitudes in the $KK'$-basis, we need to specify the lead states. We assume that the lead states constitute Kramers doublets, which we will denote:
\begin{equation}\label{lskd}
\ket{\alpha\nu\up}=T\ket{\alpha\tilde{\nu}\down},
\end{equation}
where $\alpha=L,R$ is a lead index. We also assume that the Hamiltonian describing electrons in the leads is real and that we can choose arbitrary spin-quantization direction (for example, there is no spin-orbit coupling in the leads). In the case of no spin-orbit interaction the states (\ref{lskd}) can be represented in real space as
\begin{equation}\label{replskd}
\braket{\br}{\alpha\nu\up}=
\begin{pmatrix}
a_{\alpha\nu}(\br) \\ 0
\end{pmatrix}, \quad
\braket{\br}{\alpha\tilde{\nu}\down}=
\begin{pmatrix}
0 \\ a_{\alpha\nu}(\br)
\end{pmatrix},
\end{equation}
where $a_{\alpha\nu}(\br)$ is a real function of coordinates. Here the spin-quantization direction is chosen along the tube axis. When there is no magnetic field, we can define the following tunneling amplitudes
\begin{equation}\label{eota}
\begin{aligned}
&t_{\alpha\nu\up}^{K\up}=\bra{\alpha\nu\up}H_{\mathrm{tot}}\ket{K\up}=t_{\alpha\nu,1},\\
&t_{\alpha\nu\down}^{K'\down}=\bra{\alpha\tilde{\nu}\down}H_{\mathrm{tot}}\ket{K'\down}=t_{\alpha\nu,1}^{*},\\
&t_{\alpha\nu\down}^{K\down}=\bra{\alpha\tilde{\nu}\down}H_{\mathrm{tot}}\ket{K\down}=t_{\alpha\nu,2},\\
&t_{\alpha\nu\up}^{K'\up}=\bra{\alpha\nu\up}H_{\mathrm{tot}}\ket{K'\up}=t_{\alpha\nu,2}^{*},
\end{aligned}
\end{equation}
where $H_{\mathrm{tot}}$ is the total single particle Hamiltonian representing leads, dot region, and tunneling barriers. We also used time-reversal symmetry of the states. It can be shown (using a real space representation of carbon nanotube states described in Refs.~\onlinecite{ando2005,bula2008,weiss2010}) that the difference between tunneling amplitudes $t_{\alpha\nu,1}$ and $t_{\alpha\nu,2}$ appears due to spin-orbit coupling and that it can be neglected, when spin-orbit coupling is small compared to curvature induced splitting in the carbon nanotube (see \appendixname~\ref{sec:appendixA}), i.e.,
\begin{equation}
t_{\alpha\nu,1}\approx t_{\alpha\nu,2} \quad\rightarrow\quad  t_{\alpha\nu}.
\end{equation}
However, if the states in the leads cannot be chosen to be real (for instance, if there is spin-orbit coupling in the leads), then the two Kramers doublets can be coupled differently ($t_{\alpha\nu,1}\neq t_{\alpha\nu,2}$), which we will also consider. Note that in general if there is spin-orbit coupling in the leads, the lead states have the form
\begin{equation}\label{replskd2}
\braket{\br}{\alpha\nu\up}=
\begin{pmatrix}
a_{\alpha\nu}(\br) \\ b_{\alpha\nu}(\br)
\end{pmatrix}, \quad
\braket{\br}{\alpha\tilde{\nu}\down}=
\begin{pmatrix}
-b^{*}_{\alpha\nu}(\br) \\ a^{*}_{\alpha\nu}(\br)
\end{pmatrix},
\end{equation}
where $a_{\alpha\nu}(\br)$, $b_{\alpha\nu}(\br)$ are complex functions of coordinate. This additionally may introduce non-vanishing tunneling amplitudes
\begin{equation}\label{eota2}
\begin{aligned}
&t_{\alpha\tilde{\nu}\down}^{K\up}=\bra{\alpha\tilde{\nu}\down}H_{\mathrm{tot}}\ket{K\up}=t_{\alpha\nu,3},\\
&t_{\alpha\nu\up}^{K'\down}=\bra{\alpha\nu\up}H_{\mathrm{tot}}\ket{K'\down}=-t_{\alpha\nu,3}^{*},\\
&t_{\alpha\nu\up}^{K\down}=\bra{\alpha\nu\up}H_{\mathrm{tot}}\ket{K\down}=t_{\alpha\nu,4},\\
&t_{\alpha\tilde{\nu}\down}^{K'\up}=\bra{\alpha\tilde{\nu}\down}H_{\mathrm{tot}}\ket{K'\up}=-t_{\alpha\nu,4}^{*}.
\end{aligned}
\end{equation}
The inclusion of such terms does not change the qualitative picture of the results presented in the article, so for simplicity we will neglect them.

In general the tunneling amplitudes (\ref{eota}) also depend on parallel magnetic field, however this effect is small, when magnetic field is small compared to curvature induced splitting, and we do not consider it in our calculations (see \appendixname~\ref{sec:appendixA}). We also neglect $\nu$ dependence of the tunneling amplitudes $t_{\alpha\nu,1/2}\rightarrow t_{\alpha,1/2}$.

\section{\label{sec:s3}Tunneling induced level shifts}

We want to determine how the tunneling renormalizes the energy of the many-body states of Hamiltonian $H_{\LR}+H_{\D}$ in a subset $\ket{m}\in\{\ket{\m{D}}\ket{\LR}\}$ using quasi-degenerate perturbation theory (``L{\"o}wdin partitioning'').\cite{lowdin1951,wink2003} Here state $\ket{\m{D}}$ is one of the sixteen many-body states of the dot Hamiltonian $H_{\D}$, and $\ket{\LR}$ is the zero temperature ground state of the leads Hamiltonian $H_{\LR}$. We perform a unitary transformation $\e^{\i S}$ to Hamiltonian (\ref{ham}) in such a way that for the transformed Hamiltonian
\begin{equation}\label{effham}
\tilde{H}=\e^{-\i S}H\e^{\i S},
\end{equation}
the matrix elements $\bra{m}\tilde{H}\ket{l}$ vanish to the desired order in $H_{\T}$, where $\ket{l}\in\{\ket{\m{D}}\ket{\LR'}\}$ are states with a different lead state of $H_{\LR}$ than the ground state ($\ket{\LR'}\neq\ket{\LR}$). The matrix elements of effective Hamiltonian (\ref{effham}) for $\ket{m}$ states take the following form
\begin{equation}
\tilde{H}_{mm'}= H^{(0)}_{mm'}+H^{(2)}_{mm'}+H^{(4)}_{mm'}+\ldots,
\end{equation}
where odd powers of expansion in $H_{\T}$ vanish. We specify only the zeroth- and second-order expressions\cite{wink2003}
\begin{subequations}
\begin{align}
&\label{pe0}H^{(0)}_{mm'}=\left(H^{\LR}_{mm}+H^{\D}_{mm}\right)\delta_{mm'},\\
&\label{pe2}H^{(2)}_{mm'}=\frac{1}{2}\sum_l H^{\T}_{ml}H^{\T}_{lm'}
\left(\frac{1}{E_m-E_l}+\frac{1}{E_{m'}-E_l}\right),
\end{align}
\end{subequations}
where $E$ are energies of the states. We also have changed the subscripts $\D$, $\LR$, $\T$ into superscripts for convenience.

Expression (\ref{pe0}) simply gives the energy of the state $\ket{m}$ if $m=m'$. In order to evaluate expressions (\ref{pe2}), we assume that tunneling amplitudes do not depend on $\nu$ ($t_{\alpha\nu \s}^{n}\rightarrow t_{\alpha\s}^{n}$) and will perform $\nu$-sums in expression (\ref{pe2}) using the flat band approximation for the leads spectrum, i.e.,
\begin{equation}
\sum_{\nu}t_{\alpha\nu \s}^{n}\ldots\rightarrow\rho_{\alpha\s}t_{\alpha\s}^{n}\int_{-D}^{D} d\xi\dots
\end{equation}
where $\xi=\ve_{\alpha\nu}-\mu_{\alpha}\in[-D\ldots D]$ corresponds to the flat band. Here $\rho_{\alpha\s}$ is the density of states at the Fermi energy for electrons in the lead $\alpha$ with spin $\s$. We also assume that the bandwidth $D$ is much larger than other energy scales in the problem. After performing the sum over intermediate lead states $\ket{\LR'}$ and also the $\nu$-sums, we get the following expression for the second-order matrix element between states $\ket{\m{D}}\ket{\LR}$ and $\ket{\m{D}'}\ket{\LR}$:
\begin{widetext}
\begin{equation}
\begin{aligned}\label{soehm}
H^{(2)}_{\m{D}\m{D'}}
\approx \frac{1}{2}\sum_{\substack{\alpha\s\\nn' \\ \ket{\psi_{\D}} }}
\rho_{\alpha\s}
&\Biggr[
t_{\alpha\s}^{ n}(t_{\alpha\s}^{n'})^{*}
\bra{\m{D}}\dan_{n}\ket{\psi_{\D}}\bra{\psi_{\D}}\dd_{n'}\ket{\m{D'}}
\left(\ln\absB{\frac{E_{\m{D}}-E_{\psi_{\D}}+\mu_{\alpha}}{D}}
+\ln\absB{\frac{E_{\m{D'}}-E_{\psi_{\D}}+\mu_{\alpha}}{D}}\right)\\
&+(t_{\alpha\s}^{n})^{*}t_{\alpha\s}^{n'}
\bra{\m{D}}\dd_{n}\ket{\psi_{\D}}\bra{\psi_{\D}}\dan_{n'}\ket{\m{D'}}
\left(\ln\absB{\frac{E_{\m{D}}-E_{\psi_{\D}}-\mu_{\alpha}}{D}}
+\ln\absB{\frac{E_{\m{D'}}-E_{\psi_{\D}}-\mu_{\alpha}}{D}}\right)\Biggr].
\end{aligned}
\end{equation}
\end{widetext}
Here the sum runs over all possible many-body states $\ket{\psi_{\D}}$ of the dot Hamiltonian $H_{\D}$. More detailed derivation of Eq.~(\ref{soehm}) is given in \appendixname~\ref{sec:appendixR}.

We see that the effective Hamiltonian $\tilde{H}\approx H^{(0)}+H^{(2)}$ has a block structure to second order in $H_{\T}$, i.e., only the matrix elements between the same charge dot states $\ket{\m{D}}=\ket{Nl}$ and $\ket{\m{D'}}=\ket{Nl'}$ are non-zero, i.e., $H_{Nl,N'l'}=H_{Nl,Nl'}\delta_{NN'}$. Here we chose to classify the states by the charge number $N$ and quantum number $l$. After diagonalizing the effective $16\times16$ Hamiltonian $\tilde{H}$, we find the new eigenspectrum $\tilde{E}$ to second order in $H_{\T}$. In the next section, we will examine the cotunneling thresholds in the middle of the single charge diamond. We will be interested in the regime where the charging energy is much larger than the single particle spectrum energies $\ve_n$, bias $V$, and tunneling rates $\Gamma_{\alpha\s}^{n}=\pi\rho_{\alpha\s}\abs{t_{\alpha\s}^{n}}^2$, in order for our expansion in $H_{\T}$ to be valid. Note that all energies are measured in units of $\Gamma=\sum_{\alpha j}\Gamma_{\alpha, j}=\pi\rho\sum_{\alpha j}\abs{t_{\alpha, j}}^2$, where $t_{\alpha j}$ are tunneling amplitudes defined by (\ref{eota}), with neglected $\nu$ dependence, and assuming spin and lead independent density of states throughout, i.e., $\rho_{\alpha s}=\rho$. In \appendixname~\ref{sec:appendixR}, we also consider fourth-order energy shifts for the cases discussed in Sections \ref{sec:s4} and \ref{sec:s5}, in order to show the validity of the expansion. For the parameter regime discussed in the paper, we confirm that fourth-order corrections to the energy shifts are irrelevant near the middle of the diamond.

\section{\label{sec:s4}Tunneling renormalized cotunneling}

\begin{figure*}[ht]
\begin{center}
\includegraphics[width=0.8\textwidth]{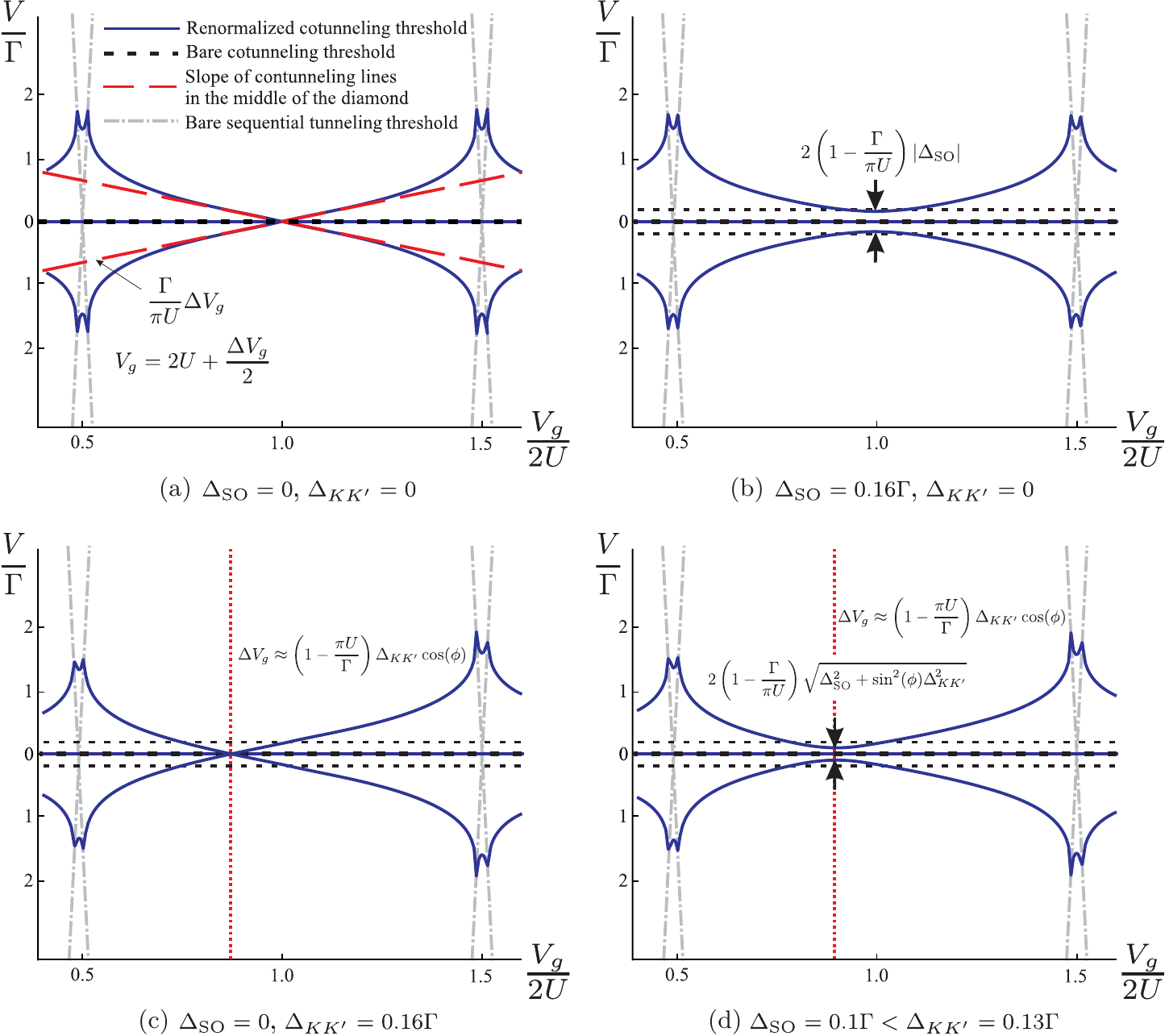}
\caption{\label{fig2}(Color online) Tunneling renormalized cotunneling thresholds shown as solid curves (blue) for different $KK'$-mixing and spin-orbit coupling values. The dashed lines (black) depict bare cotunneling thresholds and dashed-dotted lines (gray) show bare sequential tunneling thresholds. The long dashed lines (red) in a) show the linearized cotunneling threshold, when there is no $KK'$-mixing and spin-orbit coupling, and the dotted lines (red) in c) and d) depict, respectively, the position of zero bias crossing and minimum separation of the cotunneling threshold from zero bias. The values of the other parameters are $B=0$, $U=32\Gamma$, $D=10^9\Gamma$, with all tunneling amplitudes $t_{L/R,1/2}$ being equal and $\Arg[t_{L/R,1/2}]=0$.}
\end{center}
\end{figure*}

In the Coulomb blockade regime with a particular charge state $N$ of the dot, an inelastic cotunneling process occurs when the bias voltage $V$ matches the energy difference between two different orbital states of that charge state ($\abs{V_{\mathrm{th}}}=E_{Nl}-E_{Nl'}$) and a threshold for charge transport from the source to the drain electrode through the quantum dot is reached. We consider cotunneling processes involving the ground state $E_{Nl'}$ and some excited state $E_{Nl}$. In this section, we are interested in how this cotunneling threshold is modified by the tunneling in the middle of the single charge $N=1$ diamond ($V_g=2U+\Delta V_g/2$). We solve the equation
\begin{equation}
\abs{\tilde{V}_{\mathrm{th}}}=\Delta E(\bB,V_g,\tilde{V}_{\mathrm{th}})=\tilde{E}_{Nl}-\tilde{E}_{Nl'},
\end{equation}
where $\tilde{E}$ is the tunneling renormalized eigenspectrum.

First we study the case when $B=0$, all single particle orbitals are coupled symmetrically to the leads $\Gamma_{\alpha,j}=\Gamma/4$, and the tunneling rate $\Gamma$ is much larger than the disorder splitting $\Delta_{KK'}\approx0$ and spin-orbit coupling $\Delta_{\SO}\approx0$, with the resulting tunneling renormalized threshold shown in \figurename~\ref{fig2}a. We see that the fourfold degenerate ``shell'' spectrum is split in a gate dependent way, and we get the $1^{\mathrm{st}}$ cotunneling line at zero bias (for $B\neq0$ it is visible at finite bias),\cite{[{For $T<T_K$ it evolves into a Kondo peak, which is not included in our approach}] empty} and one twofold degenerate gate dependent line (which splits into the $2^{\mathrm{nd}}$ and $3^{\mathrm{rd}}$ cotunneling lines for $B\neq 0$). This splitting appears due to effective mixing of $K$ and $K'$ states because of tunneling to the leads. We can see this more clearly by examining the single $N=1$ charge $4\times4$ block of the effective Hamiltonian (\ref{soehm}), which in this case is
\begin{equation}\label{fiwhs}
\m{H}=
\bordermatrix{&\ket{1,K\up}   &\ket{1,K'\down} &\ket{1,K\down} &\ket{1,K'\up} \cr
                &\m{H}_{d}    &0             &0            &\m{H}_{o}\cr
                &0            &\m{H}_{d}     &\m{H}_{o} &0   \cr
                &0            &\m{H}_{o}  &\m{H}_{d}    &0   \cr
                &\m{H}_{o} &0             &0            &\m{H}_{d}   },
\end{equation}
where $\ket{1,l}$ denotes the single charge many-body eigenstates of Hamiltonian (\ref{hamD}), with $l$ being the occupied single particle state.
The off-diagonal term is
\begin{equation}\label{todtfss}
\m{H}_{o}=\frac{\Gamma}{4\pi}\sum_{\alpha}
\ln\absB{\frac{1+\frac{\Delta V_g+\alpha V}{2U}}{1-\frac{\Delta V_g+\alpha V}{2U}}}\approx\frac{\Gamma}{\pi}\frac{\Delta{V}_g}{2U},
\end{equation}
and $\m{H}_{d}$ is the diagonal term, which is the same for all single charge states. Also here $\alpha=+1(-1)$ for the left (right) lead, and in Eq. (\ref{todtfss}) we have linearized the logarithms in $\tfrac{\Delta{V}_g+\alpha V}{2U}$. The Hamiltonian (\ref{fiwhs}) has the following two-fold degenerate eigenvalues
\begin{equation}\label{scseskpm}
E_{K^{\pm}s}=\m{H}_d\pm\frac{\Gamma}{\pi}\frac{\Delta{V}_g}{2U}, \quad s=\up,\down,
\end{equation}
with corresponding eigenstates
\begin{equation}
\ket{1,K^{\pm}\s}=\frac{1}{\sqrt{2}}\left(\ket{1,K\s}\pm\ket{1,K'\s}\right).
\end{equation}
On the left side of the diamond ($\Delta{V}_g<0$) the ground states are the states $\ket{1,K^+\s}$ and on the right side ($\Delta{V}_g>0$) are the states $\ket{1,K^-\s}$. So qualitatively \figurename~\ref{fig0}b represents the result in \figurename~\ref{fig2}a. From (\ref{scseskpm}) eigenspectrum we find that the slope of the $2^{\mathrm{nd}}$, $3^{\mathrm{rd}}$ cotunneling lines is given by (dashed (red) line in \figurename~\ref{fig2}a)
\begin{equation}\label{soscl}
S=\pm\frac{\Gamma}{\pi U}\Delta V_g,
\end{equation}
We see that there is a crossing of cotunneling lines exactly in the middle of the diamond, i.e., the tunneling renormalization effects vanish. If the tunneling couplings are different for the left and the right leads, the positive and the negative slopes get corrections and instead of the above expression (\ref{soscl}), we obtain for the slope
\begin{equation}
S_{\pm}=\pm\frac{\Gamma_L+\Gamma_R}{\pi U}\left(1\pm\frac{\Gamma_L-\Gamma_R}{\pi U}\right)\Delta V_g,
\end{equation}
when the tunneling couplings are real, and where $\Gamma_{\alpha}=\Gamma_{\alpha1}+\Gamma_{\alpha2}$. We see that asymmetry between positive and negative bias appears, however, this corresponds only to second-order effect in $\Gamma/U$.

When spin-orbit coupling is included, instead of cotunneling lines crossing in the middle of the diamond, we get an anticrossing of size (\figurename~\ref{fig2}b)
\begin{equation}
A_1=2\left(1-\frac{\Gamma}{\pi U}\right)\abs{\Delta_{\SO}}.
\end{equation}
\begin{figure}[ht]
\begin{center}
\includegraphics[width=0.8\columnwidth]{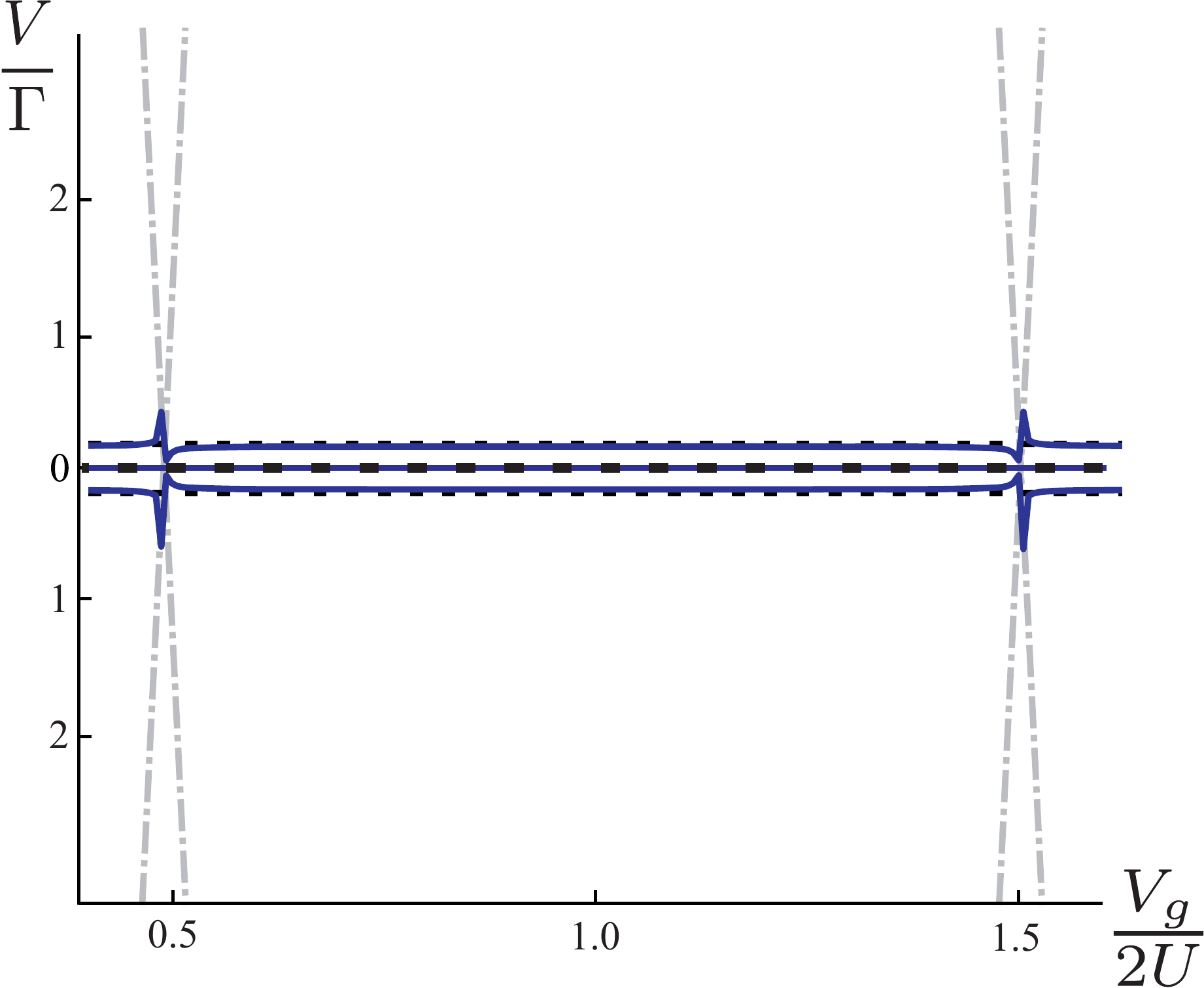}
\caption{\label{fig3}(Color online) Illustration of tunneling renormalization reduction, when there is a phase difference between the left and right lead tunneling couplings, and the relations (\ref{cc}) are satisfied. The values of parameters are $\Delta_{\SO}=0.1\Gamma$, $\Delta_{KK'}=0.13\Gamma$, $B=0$, $U=32\Gamma$, $D=10^9\Gamma$, $\tfrac{t_{R,1/2}}{t_{L,1/2}}=\e^{\i\pi/2}$, $\Arg[t_{L,1/2}]=0$. The legend is the same as in \figurename~\ref{fig2}.}
\end{center}
\end{figure}
It is not possible to restore a crossing if both Kramers doublets defined in \sectionname~\ref{sec:modelA} have the same tunneling couplings to the leads ($t_{\alpha,1}\approx t_{\alpha,2}$), even if mixing of $K$ and $K'$ due to disorder $\Delta_{KK'}$ is included (\figurename~\ref{fig2}d). In this case, for symmetric couplings to the left and right leads we can find the eigenspectrum around the middle of the diamond by linearizing the logarithms in Eq.~(\ref{soehm}), which yields the following energy difference between ground and excited states
\begin{equation}\label{edwiae}
\begin{aligned}
\Delta E=&\Bigg\{\left[\left(1-\frac{\Gamma}{\pi U}\right)\Delta_{\Sigma}+\frac{\Gamma}{\pi U}\cos(\phi)\frac{\Delta_{KK'}}{\Delta_{\Sigma}}\Delta{V}_g\right]^2\\
&+\absB{\frac{\Gamma}{\pi U}\left(\frac{\Delta_{\SO}}{\Delta_{\Sigma}}\cos(\phi)+\i\sin(\phi)\right)\Delta{V}_g}^2\Bigg\}^{1/2},
\end{aligned}
\end{equation}
where
\begin{equation}
\Delta_{\Sigma}=\sqrt{\Delta_{\SO}^2+\Delta_{KK'}^2},
\end{equation}
and the phase $\phi$ is the sum of the $KK'$-mixing phase ($\Delta_{KK'}=\abs{\Delta_{KK'}}\e^{\i\phi_{KK'}}$) and the tunneling amplitudes phase ($t_{L/R,1/2}=\abs{t}\e^{\i\phi_t}$):
\begin{equation}
\phi=\phi_{KK'}+2\phi_{t}.
\end{equation}
From the above expression (\ref{edwiae}) we find that an anticrossing appears near the point
\begin{equation}\label{acwskkp}
\Delta{V}_g\approx\left(1-\frac{\pi U}{\Gamma}\right)\Delta_{KK'}\cos(\phi),
\end{equation}
and its size is
\begin{equation}
A_2=2\left(1-\frac{\Gamma}{\pi U}\right)\sqrt{\Delta_{\SO}^2+\sin^2(\phi)\Delta_{KK'}^2}.
\end{equation}
We note that in this case the middle of the bare diamond is given by $\Delta{V}_g=-\Delta_{\Sigma}$. 
 If spin-orbit coupling is neglected $\Delta_{\SO}=0$, but there is $KK'$-mixing $\Delta_{KK'}$, and the total phase $\phi$ is equal to zero, a crossing instead of anticrossing appears near the point (\ref{acwskkp}), as shown in \figurename~\ref{fig2}c.

By changing the relative phases between the left and right couplings, we can reduce the tunneling renormalization, as shown in \figurename~\ref{fig3}. The condition for complete reduction of the tunneling renormalization around the middle of the diamond is
\begin{subequations}\label{cc}
\begin{align}
&\label{cc1}t_{L,1}t_{L,2}+t_{R,1}t_{R,2}=0,\\
&\label{cc2}\abs{t_{L,1}}^2+\abs{t_{R,1}}^2=\abs{t_{L,2}}^2+\abs{t_{R,2}}^2,
\end{align}
\end{subequations}
which can be rewritten as
\begin{equation}
t_{R,1}=\e^{\i\vphi}t_{L,2}, \quad t_{R,2}=-\e^{-i\vphi}t_{L,1},
\end{equation}
where $t_{L/R,1/2}$ are complex numbers, and $\vphi$ is some arbitrary phase.
\begin{figure}[ht]
\begin{center}
\includegraphics[width=0.8\columnwidth]{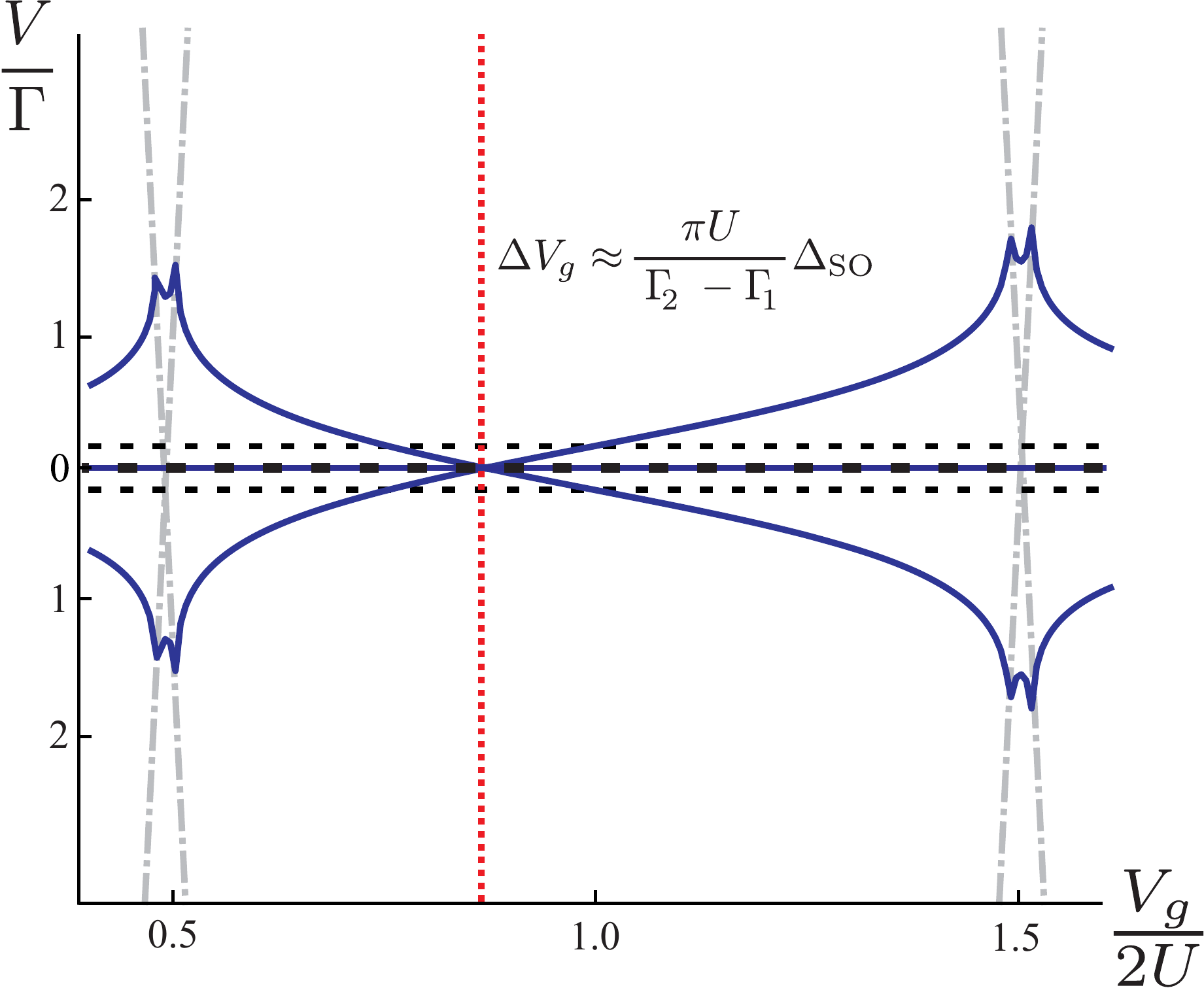}
\caption{\label{fig4}(Color online) Illustration of a zero bias crossing, when there is only spin-orbit coupling, and the Kramers doublets have different tunneling couplings to the leads (only relation (\ref{cc1}) is satisfied). The values of the parameters are $\Delta_{\SO}=0.16\Gamma$, $\Delta_{KK'}=B=0$, $U=32\Gamma$, $D=10^9\Gamma$, $\absB{\tfrac{t_{L/R,1}}{t_{L/R,2}}}=7$, $\Arg[t_{L,1/2}]=\Arg[t_{R1}]=0$, $\Arg[t_{R2}]=\pi$. The legend is the same as in \figurename~\ref{fig2}.}
\end{center}
\end{figure}

It is possible to get a crossing if the Kramers doublets are coupled differently, i.e., $t_{\alpha, 1}\neq t_{\alpha, 2}$, as shown in \figurename~\ref{fig4}, where we for simplicity consider the case with only spin-orbit coupling, but the statement is also true when $KK'$-mixing is included. In this case, the condition for crossing is given by relation (\ref{cc1}), when relation (\ref{cc2}) is not satisfied. Then the position of the crossing is given by
\begin{equation}\label{efcc}
\Delta V_g\approx\frac{\pi U}{\Gamma_2-\Gamma_1}\Delta_{\mathrm{SO}},
\end{equation}
where $\Gamma_{j}=\Gamma_{L,j}+\Gamma_{R,j}$, and the middle of the bare diamond in this case is $\Delta{V}_g=-\abs{\Delta_{\SO}}$. To be able to observe it experimentally, the value of Eq.~(\ref{efcc}) has to be between $-2U$ and $2U$.

\section{\label{sec:s5}Gate dependence of $g$-factors}

\begin{figure*}[ht]
\begin{center}
\includegraphics[width=0.8\textwidth]{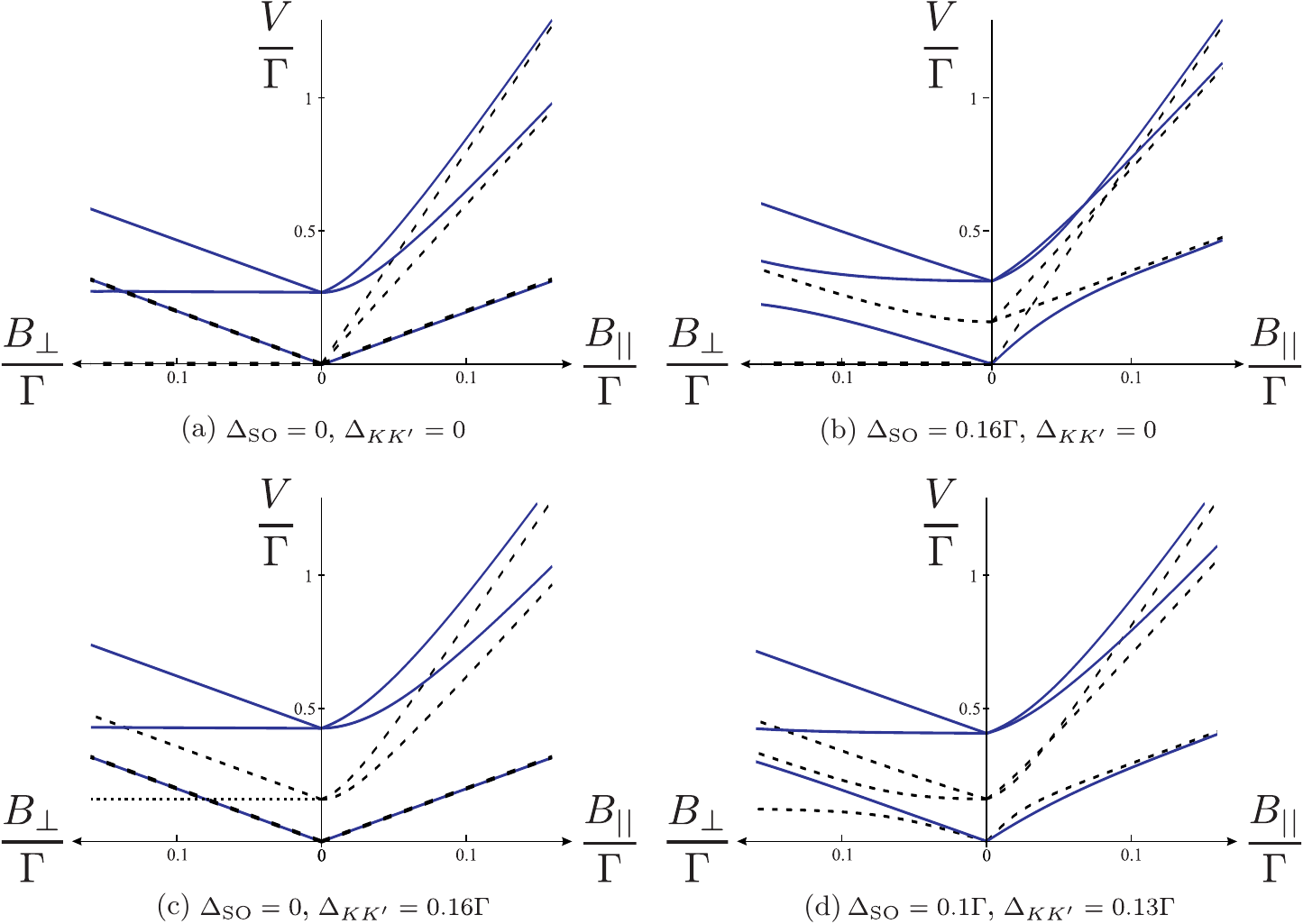}
\caption{\label{fig5} (Color online) Comparison of the dependence on parallel $B_{||}$ and perpendicular $B_{\perp}$ magnetic fields of the bare (dashed curves (black)) and the tunneling renormalized (solid curves (blue)) cotunneling thresholds for different values of $KK'$-mixing and spin-orbit coupling at a gate voltage $V_g=1.2\times2U$ away from the middle of the diamond. The values of the other parameters are $U=32\Gamma$, $D=10^9\Gamma$, with all tunneling amplitudes $t_{L/R,1/2}$ being equal and $\Arg[t_{L/R,1/2}]=0$.}
\end{center}
\end{figure*}

Now we investigate how the tunneling induced level shifts affect the magnetic field dependence of the cotunneling threshold, i.e., we examine gate dependence of $g$-factors. As in the previous section, we restrict our examination to the single charge $N=1$ diamond. We start by examining the case when the spin-orbit coupling and $KK'$-mixing are neglected, the magnetic field is parallel to the tube axis, and the couplings to the left and right leads are equal. When the gate voltage is exactly in the middle of the diamond the bare $g$-factors are almost unaffected and they are reduced only by a factor $(1-\Gamma/\pi U)$
\begin{equation}
\tilde{g}_{s/\orb}\approx \left(1-\kappa\right)g_{s/\orb}, \quad \kappa=\frac{\Gamma}{\pi U}.
\end{equation}
 For a very small $\Gamma/U$ ratio, the renormalized threshold matches the bare one (dashed (black) curves in \figurename~\ref{fig5}a).
For perpendicular magnetic field $B_{\perp}$ the renormalization of $g$-factors is also small and for $1^{\mathrm{st}}$, $2^{\mathrm{nd}}$, and $3^{\mathrm{rd}}$ cotunneling lines, respectively, is given by
\begin{equation}
\kappa g_s, \ (1-\kappa)g_s, \ g_s.
\end{equation}
Going away from the middle of the diamond, we find that $g$-factors acquires gate dependence (solid (blue) curves in \figurename~\ref{fig5}a), which for the $\kappa\Delta V_g>>B>0$ case is written in \tablename~I, for different transitions. The situation when spin-orbit coupling is included is depicted in \figurename~\ref{fig5}b. We see that the tunneling renormalization again acts as gate dependent $\Delta_{KK'}$ splitting. The effective $g$-factors for small magnetic $B$ fields ($\kappa \Delta V_g, \ \Delta_{\SO}>> B>0$) are written in \tablename~I, where the following notation is introduced:
\begin{subequations}
\begin{align}
\tilde{\kappa}&\approx\kappa
\left[1+\left(\frac{1-\kappa}{\kappa} \frac{\Delta_{\SO}}{\Delta{V}_g}\right)^2\right]^{-1/2},\\
\tilde{g}_s&\approx g_s
(1-\kappa)\left[1+\left(\frac{1-\kappa}{\kappa}\frac{\Delta_{\SO}}{\Delta{V}_g}\right)^2\right]^{-1/2},\\
\tilde{g}_{\orb}&\approx g_{\orb}
(1-\kappa)
\left[1+\left(\frac{\kappa}{1-\kappa} \frac{\Delta{V}_g}{\Delta_{\SO}}\right)^2\right]^{-1/2}.
\end{align}
\end{subequations}

When there is only $KK'$-mixing, the magnetic field dependence of the cotunneling threshold does not change qualitatively, and the only difference at finite $\Delta{V}_g$ is effective enhancement of the $KK'$-mixing due to tunneling-renormalization, as can be seen from \figurename~\ref{fig5}c. The situation when both spin-orbit coupling and $KK'$-mixing are included is depicted in \figurename~\ref{fig5}d.

\begin{table*}
\label{tbl1}
\begin{tabular}{|c|c|c|c|c|}\hline
Cotunneling   & \multicolumn{2}{c|}{$\kappa\Delta{V}_g>>B>0, \ \Delta_{\SO}=0$} & \multicolumn{2}{c|}{$\kappa\Delta{V}_g>\Delta_{\SO}>>B>0$} \\\cline{2-5}
line          &  Bare $g$-factor & Renormalized $g$-factor & Bare $g$-factor & Renormalized $g$-factor \\\hline
\multicolumn{5}{c}{Parallel field $B_{||}$}     \\\hline
$1^\text{st}$ & $g_s$           & $\left(1-\kappa\right)g_s$                           & $g_s+g_{\orb}$ & $(1-\kappa-\tilde{\kappa})g_s+\tilde{g}_{\orb}$\\\hline
$2^\text{nd}$ & $g_{\orb}$      & $\kappa g_s+\frac{g_{\orb}^2B}{2\kappa \Delta{V}_g}$ & $g_s$          & $(1-\kappa)g_s$  \\\hline
$3^\text{rd}$ & $g_s+g_{\orb}$  & $g_s+\frac{g_{\orb}^2B}{2\kappa \Delta{V}_g}$        & $g_{\orb}$     & $-\tilde{\kappa}g_s+\tilde{g}_{\orb}$ \\\hline
\multicolumn{5}{c}{Perpendicular field $B_{\perp}$}                                      \\\hline
$1^\text{st}$ & $0$         & $g_s$           & $0$                           & $\kappa g_s+\tilde{g}_s$ \\\hline
$2^\text{nd}$ & $g_s$       & $\kappa g_s$    & $\frac{g_s B}{2\Delta_{\SO}}$ & $\kappa g_s$ \\\hline
$3^\text{rd}$ & $g_s$       & $(1-\kappa)g_s$ & $\frac{g_s B}{2\Delta_{\SO}}$ & $\tilde{g}_s$ \\\hline
\end{tabular}
\caption{Bare and renormalized $g$-factors for the carbon nanotube quantum dot, when $g_{\orb}>g_s$ and $\Delta_{KK'}=0$.}
\end{table*}

\begin{figure}[ht]
\begin{center}
\includegraphics[width=0.8\columnwidth]{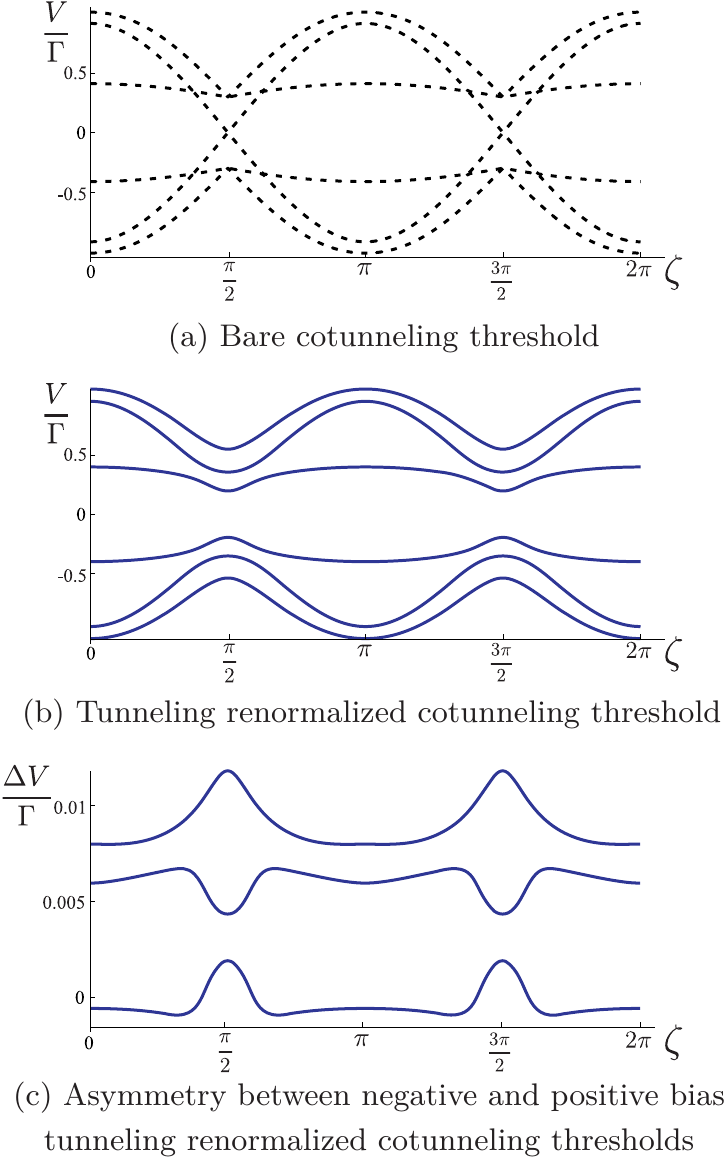}
\caption{(Color online) Angle dependence of the cotunneling threshold. The values of the parameters are $B=0.13\Gamma$, $\Delta_{\SO}=0.16\Gamma$, $\Delta_{KK'}=0$, $U=32\Gamma$, $D=10^9\Gamma$, $\tfrac{t_{L,1/2}}{t_{R,1/2}}=7$, $\Arg[t_{L/R,1/2}]=0$.}
\label{fig6}
\end{center}
\end{figure}

In an experiment, the cotunneling threshold dependence on the magnetic field angle $\zeta$ (with respect to the tube axis) also could be measured. The angle dependence of the bare cotunneling threshold is shown in \figurename~\ref{fig6}a. Again, in the middle of the diamond there is almost no renormalization due to tunneling and it matches \figurename~\ref{fig6}a, and the situation at finite $\Delta{V_g}$ is shown in \figurename~\ref{fig6}b. If the quantum dot is coupled to the leads asymmetrically, then the asymmetry between positive and negative bias thresholds acquires angle dependence, which can be seen by adding positive and negative bias thresholds in \figurename~\ref{fig6}b, and the result is shown in \figurename~\ref{fig6}c.

\section{\label{sec:concl}Conclusions}

In this paper we have examined tunneling renormalization of the quantum dot cotunneling spectrum by considering energy shifts of many-body eigenstates using quasi-degenerate perturbation theory\cite{lowdin1951,wink2003} in tunneling Hamiltonian $H_{\T}$. The second-order result Eq.~(\ref{soehm}) is applicable to any quantum dot with an arbitrary number of single particle orbitals, when the tunneling rates $\Gamma$ are much smaller than half of the charging energy ($\Gamma<<U$), and gate voltage is far from the charge degeneracy points.

Using this second-order result we determined the energy shifts for the carbon nanotube quantum dot, where a fourfold ``shell'' structure (\sectionname~\ref{sec:modelA}) of the single particle spectrum was assumed. It was shown that tunneling renormalization introduces gate dependent $KK'$-mixing of carbon nanotube orbitals, and this in turn renormalizes $g$-factors for some cotunneling lines in a gate dependent way. From the energy shifts the cotunneling spectrum was obtained and we found that for asymmetric tunneling couplings to the right and left leads bias asymmetry appears, which is a second-order effect in small parameter $\Gamma/U$. By measuring the cotunneling threshold asymmetry (if the coupling to the left and to the right lead is different) between positive and negative bias and its dependence on magnetic field angle with respect to the tube axis, it would be possible to indicate that the gate dependence of cotunneling lines appears due to tunneling renormalization and not some other effects, e.~g., a change of the local electrostatic potential. It was also found that the tunneling renormalization can be reduced by changing the relative phases between the left and right couplings to the leads.

\begin{acknowledgments}
We thank Kasper Grove-Rasmussen, Jonas Hauptmann, and Martin Leijnse for valuable discussions. The research leading to these results has received funding from the European Union Seventh Framework Programme (FP7/2007-2013) under the grant agreement no 270369 (ELFOS).
\end{acknowledgments}

\appendix

\section{\label{sec:appendixA} $KK'$-basis states and tunneling amplitudes}

In this \appendixname, we discuss the explicit form of $KK'$-basis states for a single ``shell'' used in Hamiltonian (\ref{kkphm}), and corresponding tunneling amplitudes (\ref{eota}) and (\ref{eota2}) to these states. The effective single particle Hamiltonian for the carbon nanotube derived using $\bk\cdot\bp$ expansion near so-called $K$ and $K'$ points in the first Brillouin zone is given in sublattice $\sigma$ space by \cite{ando2005,izum2009}
\begin{equation}\label{finham2}
H=\hbar v_F
\begin{pmatrix}
\tau sk_0   &\tilde{k}_c-\i\tilde{k}_t \\\
\tilde{k}_c+\i\tilde{k}_t & \tau sk_0
\end{pmatrix},
\end{equation}
where
\begin{subequations}
\begin{align}
&\tilde{k}_c=\tau (k_c+k_{\Phi})-k_{c,\mathrm{cv}}-\tau s k_{\mathrm{SO}},\\
&\tilde{k}_t=k_t-\tau k_{t,\mathrm{cv}},\\
&k_c=\left(m-\frac{\nu\tau}{3}\right)\frac{1}{R}, \quad m\in \mathbb{Z}, \quad \nu=0,\pm1,
\end{align}
\end{subequations}
with $\tau=+1(-1)$ for $K(K')$, $s=+1(-1)$ for spin up (down), the above value of $\nu$ depending on the type of the nanotube, $R$ denoting its radius, and $v_F\approx8.1\times10^{14} \ \mathrm{nm/s}$ being Fermi velocity in a nanotube. Also $k_c$ corresponds to wave number along the circumferential direction of the tube, which we will denote $c$, and $k_t$ along the tube axis direction, which we will denote $t$. The other parameters are given by
\begin{subequations}
\begin{align}
&k_{\mathrm{SO}}=\alpha_1\frac{\Delta_C}{2\hbar v_F R}, &&\alpha_1=0.048 \ \mathrm{nm},\\
&k_0=\alpha_2\frac{\Delta_C}{2\hbar v_FR}\cos(3\theta), &&\alpha_2=-0.045 \ \mathrm{nm}, \\
&k_{c,\mathrm{cv}}=\beta\frac{\cos(3\theta)}{4\hbar v_FR^2}, &&\beta=24 \ \mathrm{meV\times nm^2}, \\
&k_{t,\mathrm{cv}}=\zeta\frac{\sin(3\theta)}{4R^2}, &&\zeta=-0.18 \ \mathrm{nm},\\
&k_{\Phi}=\frac{\Phi_{AB}}{\Phi_0}\frac{1}{R}=\frac{\pi e}{h}RB_{||}, &&\Phi_{AB}=\pi R^2B_{||}, \ \Phi_{0}=\frac{h}{e},
\end{align}
\end{subequations}
where the terms $k_0$, $k_{\SO}$ are due to hybridization of so-called $\sigma-\pi$ bands induced by spin-orbit coupling, $k_{c,\mathrm{cv}}$, $k_{t,\mathrm{cv}}$ are induced by curvature of the tube, and $k_{\Phi}$ is due to Aharanov-Bohm flux $\Phi_{AB}$ through the cross section of the tube. Also $\theta\in[0,\pi/6]$ denotes the chiral angle of the tube, $\Delta_{C}\approx6 \ \mathrm{meV}$ is the strength of the atomic spin-orbit coupling, and $\alpha_1$, $\alpha_2$, $\beta$, $\zeta$ are constants calculated using the tight-binding approach.\cite{izum2009} If the carbon nanotube has a finite length $L$ and is confined by very sharp rectangular potential near the ends, then the wavevector $k_t$ along the tube axis becomes discrete\cite{bula2008,jesp2011}
\begin{equation}
k_{t}\approx \frac{n\pi}{L}, \ n\in\mathbb{Z}.
\end{equation}

The eigenvalues of Hamiltonian (\ref{finham2}) are
\begin{equation}\label{eeofh}
E_{\tau,s,k_c,k_t}=\hbar v_F\left(\pm\sqrt{\tilde{k}_c^2+\tilde{k}_t^2}+\tau sk_0\right),
\end{equation}
and eigenstates consistent with $\bk\cdot\bp$ expansion in sublattice $\sigma$ space are
\begin{equation}\label{wfocnt}
\begin{aligned}
&\Psi_{\tau,s,k_c,k_t}(c,t)=\frac{1}{\sqrt{2\pi L}}\e^{\i\mathbf{K}^{\tau}\cdot \br}\e^{\i(k_c c+k_t t)}
\begin{pmatrix}
z_{\tau,s,k_c,k_t} \\ 1
\end{pmatrix}, \\
&z_{\tau,s,k_c,k_t}=\pm\frac{\tilde{k}_c-\i\tilde{k}_t}{\sqrt{\tilde{k}_c^2+\tilde{k}_t^2}},\\
&\br=\{c\cos\theta-t\sin\theta,c\sin\theta+t\cos\theta\},
\end{aligned}
\end{equation}
where the minus sign in $z_{\tau,s}$ corresponds to the valence band, the plus sign corresponds to the conduction band, and $K^{+} \ (K^{-})$ is for the $K \ (K')$ point. These are the following states, which correspond to the $KK'$-basis used in \sectionname~\ref{sec:modelA}
\begin{subequations}\label{nkokkp}
\begin{align}
\Psi_{+1,+1,k_c,k_t} \quad \rightarrow &\quad \ket{K^{\phantom{'}}\up},\\
\Psi_{-1,-1,-k_c,-k_t} \quad \rightarrow &\quad \ket{K'\down},\\
\Psi_{+1,-1,k_c,k_t} \quad \rightarrow &\quad \ket{K^{\phantom{'}}\down},\\
\Psi_{-1,+1,-k_c,-k_t} \quad \rightarrow &\quad \ket{K'\up},
\end{align}
\end{subequations}
and when there is no magnetic field they are related as
\begin{equation}\label{prekkp2}
\begin{aligned}
&\Psi_{+1,+1,k_c,k_t}=\Psi_{-1,-1,-k_c,-k_t}^{*}, \\
&\Psi_{+1,-1,k_c,k_t}=\Psi_{-1,+1,-k_c,-k_t}^{*}.
\end{aligned}
\end{equation}
We note that $-\mathbf{K}^{'}\equiv \mathbf{K}$. In the case when $\abs{k_c-\tau k_{c,\mathrm{cv}}}>>\abs{k_{\mathrm{SO}}},\abs{k_{\Phi}}$ we can expand the square root in the (\ref{eeofh}) eigenenergies to give
\begin{equation}
E_{\tau,s,k_c,k_t}\approx E_0+s\tau\frac{\Delta_{\mathrm{SO}}}{2}\mp\tau \frac{g_{\mathrm{orb}}\mu_B B_{||}}{2},
\end{equation}
where
\begin{subequations}
\begin{align}
E_0&=\pm\hbar v_F \sqrt{(\tau k_c-k_{c,\mathrm{cv}})^2+(\tau k_t-k_{t,\mathrm{cv}})^2},\\
\Delta_{\mathrm{SO}}&=
2\hbar v_F\left(k_0\pm\frac{\sgn(k_{c,\mathrm{cv}}-\tau k_c)}{\sqrt{1+\left(\frac{\tau k_t-k_{t,\mathrm{cv}}}{\tau k_c-k_{c,\mathrm{cv}}}\right)^2}}k_{\mathrm{SO}}\right),\\
g_{\mathrm{orb}}&=2
\frac{\sgn(k_{c,\mathrm{cv}}-\tau k_c)}{\sqrt{1+\left(\frac{\tau k_t-k_{t,\mathrm{cv}}}{\tau k_c- k_{c,\mathrm{cv}}}\right)^2}}
\frac{ev_FR}{2\mu_B},
\end{align}
\end{subequations}
with $\mu_B=e\hbar/2m_e$ being the Bohr magneton. Because the term $E_0$ is the same for $KK'$-basis states, after neglecting it, we arrive at the eigenspectrum (\ref{esawa}).

Now using $KK'$-basis states (\ref{nkokkp}) and lead states (\ref{replskd2}), which are written in spin $s$ space, we get the following tunneling amplitudes
\begin{align}
&\begin{aligned}
t_{\alpha\xi}^{\tau s}&=\bra{\alpha\xi}H_{\mathrm{tot}}\ket{\tau s}\\
&=(1\pm\e^{\i\vphi_{\tau,s}})\int d\br A_{\xi s}(\br)H_{\mathrm{tot}}(\br)\e^{\i \mathbf{K}\cdot\br}\e^{\i\tau (k_cc+k_tt)},
\end{aligned}\\
&\vphi_{\tau,s}=\tau\Arg\left[k_c-k_{c,\mathrm{cv}}-\i(k_t-k_{t,\mathrm{cv}})-\tau s k_{\SO}+\tau k_{\Phi}\right],\\
&A_{\xi s}=
\begin{pmatrix}
A_{\nu\up,\up} & A_{\nu\up,\down} \\
A_{\tilde{\nu}\down,\up} & A_{\tilde{\nu}\down,\down}
\end{pmatrix}
=\frac{1}{\sqrt{2\pi L}}
\begin{pmatrix}
a_{\alpha\nu}^{*} & b_{\alpha\nu}^{*} \\
-b_{\alpha\nu} & a_{\alpha\nu}
\end{pmatrix},
\end{align}
where $\xi$ denotes either $\nu\up$ or $\tilde{\nu}\down$. In general, the lead states also can depend on magnetic field $\bB$, however, we do not consider such a situation. If $\abs{k_c-k_{c,\mathrm{cv}}-\i(k_t-k_{t,\mathrm{cv}})}>>\abs{s k_{\SO}-k_{\Phi}}$, which is in most cases for carbon nanotubes, we can safely neglect the dependence of tunneling amplitudes on spin-orbit coupling ($k_{\SO}$) and parallel magnetic field ($k_{\Phi}$) to the tube axis.

\section{\label{sec:appendixB}Energy shifts with fourth-order corrections}

\begin{figure*}[ht]
\begin{center}
\includegraphics[width=0.8\textwidth]{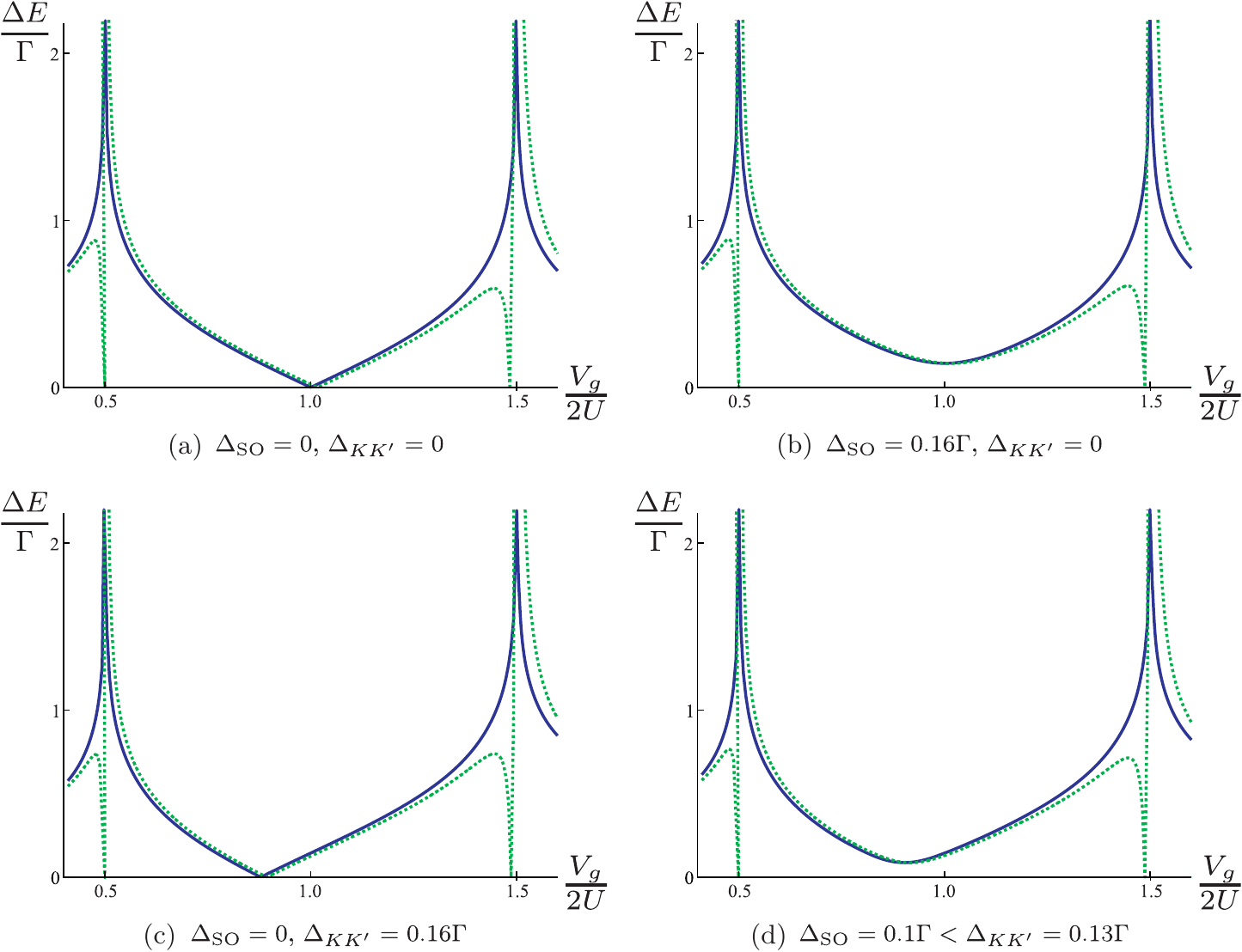}
\caption{\label{fig13}(Color online) The comparison of energy differences corresponding to cotunneling thresholds in \figurename~\ref{fig2}, where solid curves (blue) correspond to energy differences with corrections up to second order in $H_{\T}$, and dotted (green) curves correspond to energy differences with corrections up to fourth order in $H_{\T}$. The values of the other parameters for calculation are $B=0$, $U=32\Gamma$, $D=10^9\Gamma$, with all tunneling amplitudes $t_{L/R,1/2}$ being equal and $\Arg[t_{L/R,1/2}]=0$.}
\end{center}
\end{figure*}

In this \appendixname~the necessity to use larger model space than that of the single state, which is used for usual Rayleigh-Schr\"{o}dinger perturbation theory,\cite{hald1978} and appearance of off-diagonal elements in (\ref{soehm}) is discussed. Also to show the region of validity of the perturbative expansion, we consider the energy shifts with fourth-order corrections included, for the cases discussed in Sections \ref{sec:s4} and \ref{sec:s5}.

The expression for the fourth-order matrix element of the effective Hamiltonian, when projected onto the many-body states $\ket{m}$ of Eq.~(\ref{hamD}) having a particular lead state $\ket{\LR}$, is given by (\ref{foeoeh}).\cite{wink2003}
If the projection is performed to one particular many-body state $\ket{m}$, then the usual expansion of Rayleigh-Schr\"{o}dinger perturbation theory is acquired. In the case, when the single particle level spacings $\Delta$ of the quantum dot are much bigger than or comparable to the tunneling rates $\Gamma$, the expansion becomes invalid, because the first term of (\ref{foeoeh}) contains terms, which are proportional to $\Gamma\Delta^{-1}$. This situation occurs because it is possible to have intermediate states $\ket{l}$, containing the ground state of the leads $\ket{\LR}$. But, if it is projected to an extended model space, containing all many-body states having $\ket{\LR}$ ground state of the leads, this situation is resolved. However, such a procedure introduces off-diagonal elements, which also appear to second order.

The comparison of the energy differences corresponding to cotunneling thresholds in \figurename~\ref{fig2} is shown in \figurename~\ref{fig13}, where solid curves (blue) correspond to energy differences with corrections up to second order in $H_{\T}$, and dotted (green) curves correspond to energy differences with corrections up to fourth order in $H_{\T}$. From this figure we see that for chosen parameters in our calculation we have a wide range of gate voltage $V_g$ for which the second-order perturbation theory in $H_{\T}$ is valid.

\begin{widetext}
\begin{equation}
\begin{aligned}\label{foeoeh}
H_{mm'}^{(4)}&=\frac{1}{2}\sum_{l,l',l''}H^{\T}_{ml}H^{\T}_{ll'}H^{\T}_{l'l''}H^{\T}_{l''m'}
\left[\frac{1}{(E_{m}-E_{l})(E_{m}-E_{l'})(E_{m}-E_{l''})}+\frac{1}{(E_{m'}-E_{l})(E_{m'}-E_{l'})(E_{m'}-E_{l''})}\right]\\
&+\sum_{l,l',m''}H^{\T}_{ml}H^{\T}_{lm''}H^{\T}_{m''l'}H^{\T}_{l'm'}
\Biggr[\frac{8}{(E_{m}-E_{l})(E_{m}-E_{l'})(E_{m''}-E_{l'})}
+\frac{8}{(E_{m'}-E_{l})(E_{m'}-E_{l'})(E_{m''}-E_{l})}\\
&\phantom{m}+\frac{4}{(E_{m}-E_{l'})(E_{m''}-E_{l})}\left(\frac{1}{E_{m}-E_{l}}+\frac{1}{E_{m''}-E_{l'}}\right)
+\frac{4}{(E_{m'}-E_{l})(E_{m''}-E_{l'})}\left(\frac{1}{E_{m'}-E_{l'}}+\frac{1}{E_{m''}-E_{l}}\right)\\
&\phantom{m}-\frac{1}{(E_{m''}-E_{l})(E_{m''}-E_{l'})}\left(\frac{1}{E_{m}-E_{l}}+\frac{1}{E_{m'}-E_{l'}}\right)
-\frac{3}{(E_{m}-E_{l})(E_{m'}-E_{l'})}\left(\frac{1}{E_{m''}-E_{l}}+\frac{1}{E_{m''}-E_{l'}}\right)
\Biggr].
\end{aligned}
\end{equation}
\end{widetext}

\onecolumngrid

\section{\label{sec:appendixR}Explicit derivation of Eq.~(\ref{soehm})}

In this \appendixname~we present more detailed derivation of Eq.~(\ref{soehm}). After inserting the tunneling Hamiltonian (\ref{hamT}) into the second-order effective Hamiltonian expression (\ref{pe2}) and setting $m=\ket{\m{D}}\ket{\LR}$, $m'=\ket{\m{D}'}\ket{\LR}$ we obtain
\footnotesize
\begin{equation}\label{heff2}
\begin{aligned}
&\begin{aligned}
H^{(2)}_{\m{D}\m{D'},\LRo}
=\frac{1}{2}\sum_{\substack{\ket{\LR'}\neq\ket{\LR} \\ \ket{\psi_{\D}} }}
\sum_{\substack{n n'\\\alpha\nu\s \\ \alpha'\nu'\s'}}
&\left(\frac{1}{E^{{}}_{\LRo}+E^{{}}_{\m{D}}-E^{{}}_{\LR'}-E^{{}}_{\psi_{\D}}}+\frac{1}{E^{{}}_{\LRo}+E^{{}}_{\m{D'}}-E^{{}}_{\LR'}-E^{{}}_{\psi_{\D}}}\right)\\
\times\bigg[
&\bra{\LRo}\bra{\m{D}}t_{\alpha\nu\s}^{n}\cd_{\alpha\nu\s}\dan_{n}\ket{\psi_{\D}}\ket{\LR'}
\bra{\LR'}\bra{\psi_{\D}}t_{\alpha'\nu'\s'}^{n'}\cd_{\alpha'\nu'\s'}\dan_{n'}\ket{\m{D'}}\ket{\LRo}\\
+&\bra{\LRo}\bra{\m{D}}t_{\alpha\nu\s}^{n}\cd_{\alpha\nu\s}\dan_{n}\ket{\psi_{\D}}\ket{\LR'}
\bra{\LR'}\bra{\psi_{\D}}(t_{\alpha'\nu'\s'}^{n'})^{*}\dd_{n'}\can_{\alpha'\nu'\s'}\ket{\m{D'}}\ket{\LRo}\\
+&\bra{\LRo}\bra{\m{D}}(t_{\alpha\nu\s}^{n})^{*}\dd_{n}\can_{\alpha\nu\s}\ket{\psi_{\D}}\ket{\LR'}
\bra{\LR'}\bra{\psi_{\D}}t_{\alpha'\nu'\s'}^{n'}\cd_{\alpha'\nu'\s'}\dan_{n'}\ket{\m{D'}}\ket{\LRo}\\
+&\bra{\LRo}\bra{\m{D}}(t_{\alpha\nu\s}^{n})^{*}\dd_{n}\can_{\alpha\nu\s}\ket{\psi_{\D}}\ket{\LR'}
\bra{\LR'}\bra{\psi_{\D}}(t_{\alpha'\nu'\s'}^{n'})^{*}\dd_{n'}\can_{\alpha'\nu'\s'}\ket{\m{D'}}\ket{\LRo}
\bigg]
\end{aligned}\\
&\phantom{..............}\begin{aligned}
=\frac{1}{2}\sum_{\substack{\alpha\nu\s\\n n'\\\ket{\psi_{\D}}}}
\Biggr[
&t_{\alpha\nu\s}^{n}(t_{\alpha\nu\s}^{n'})^{*}
n_{\alpha\nu\s}
\bra{\m{D}}\dan_{n}\ket{\psi_{\D}}\bra{\psi_{\D}}\dd_{n'}\ket{\m{D'}}
\left(\frac{1}{\ve^{{}}_{\alpha\nu}+E^{{}}_{\m{D}}-E^{{}}_{\psi_{\D}}}+\frac{1}{\ve^{{}}_{\alpha\nu}+E^{{}}_{\m{D'}}-E^{{}}_{\psi_{\D}}}\right)\\
+&(t_{\alpha\nu\s}^{n})^{*}t_{\alpha\nu\s}^{n'}
(1-n_{\alpha\nu\s})
\bra{\m{D}}\dd_{n}\ket{\psi_{\D}}\bra{\psi_{\D}}\dan_{n'}\ket{\m{D'}}
\left(\frac{1}{-\ve^{{}}_{\alpha\nu}+E^{{}}_{\m{D}}-E^{{}}_{\psi_{\D}}}+\frac{1}{-\ve^{{}}_{\alpha\nu}+E^{{}}_{\m{D'}}-E^{{}}_{\psi_{\D}}}\right)
\Bigg],
\end{aligned}\\
\end{aligned}
\end{equation}
\normalsize
where $n_{\alpha\nu\s}=\bra{\LRo}\cd_{\alpha\nu\s}\can_{\alpha\nu\s}\ket{\LRo}\in\{0,1\}$ denotes the number of particles in the many-body state $\ket{\LRo}$ with quantum numbers $\alpha\nu\s$. After the first equality in the above expression the first and the fourth terms in the square brackets vanish, and for the second and the third term we have to satisfy $\alpha'\nu'\s'=\alpha\nu\s$, which for the second term gives $E_{\LR}-E_{\LR'}=\ve_{\alpha\nu}$ and for the third term gives $E_{\LR}-E_{\LR'}=-\ve_{\alpha\nu}$. We note that (\ref{hamLR}) corresponds to the grand-canonical Hamiltonian, and the term with the chemical potential $\mu_{\alpha}$ is only included when performing thermal averages over the lead states.

Now we will perform the thermal average of the expression (\ref{heff2}) over the lead states using the grand-canonical ensemble:
\footnotesize
\begin{equation}
\label{sorttee2}
\begin{aligned}
H^{(2)}_{\m{D}\m{D'}}=\sum_{\ket{\LRo}}W_{\LRo}^{{}}H^{(2)}_{\m{D}\m{D'},\LRo}
&=\frac{1}{2}\sum_{\substack{\alpha\nu\s\\n n'\\\ket{\psi_{\D}}}}
\Biggr[
t_{\alpha\nu\s}^{n}(t_{\alpha\nu\s}^{n'})^{*}f_{\alpha}
\bra{\m{D}}\dan_{n}\ket{\psi_{\D}}\bra{\psi_{\D}}\dd_{n'}\ket{\m{D'}}
\left(\frac{1}{\ve^{{}}_{\alpha\nu}+E^{{}}_{\m{D}}-E^{{}}_{\psi_{\D}}}+\frac{1}{\ve^{{}}_{\alpha\nu}+E^{{}}_{\m{D'}}-E^{{}}_{\psi_{\D}}}\right)\\
&+(t_{\alpha\nu\s}^{n})^{*}t_{\alpha\nu\s}^{n'}(1-f_{\alpha})
\bra{\m{D}}\dd_{n}\ket{\psi_{\D}}\bra{\psi_{\D}}\dan_{n'}\ket{\m{D'}}
\left(\frac{1}{-\ve^{{}}_{\alpha\nu}+E^{{}}_{\m{D}}-E^{{}}_{\psi_{\D}}}+\frac{1}{-\ve^{{}}_{\alpha\nu}+E^{{}}_{\m{D'}}-E^{{}}_{\psi_{\D}}}\right)
\Bigg],
\end{aligned}
\end{equation}
\normalsize
where
\begin{equation}
f_{\alpha}=f(\ve_{\alpha\nu}-\mu_{\alpha})=\frac{1}{\e^{\beta_{\alpha}(\ve_{\alpha\nu}-\mu_{\alpha})}+1}
\end{equation}
is the Fermi-Dirac distribution and
\begin{equation}
\label{gcdstb}
W_{\LRo}=\frac{1}{\m{Z}}\e^{-\beta_\mrL(E_\mrL-\mu_\mrL N_{\mrL})}\e^{-\beta_\mrR(E_\mrR-\mu_\mrR N_{\mrR})}
\end{equation}
is the probability to be in particular lead state $\ket{\LRo}$ with $\beta_{\mrL,\mrR}$ being the inverse temperatures, $N_{\mrL,\mrR}$ being the number of particles, and $E_{\mrL,\mrR}$ denoting energies of the left and the right lead, respectively. Also
\begin{equation}
\m{Z}=\sum_{\ket{\LRo}}\e^{-\beta_\mrL(E_\mrL-\mu_\mrL N_{\mrL})}\e^{-\beta_\mrR(E_\mrR-\mu_\mrR N_{\mrR})}
\end{equation}
denotes the partition function.

By assuming that the tunneling amplitudes do not depend on the quantum number $\nu$, approximating the lead eigenspectrum by a flat band (i.e., $\xi=\ve_{\alpha\nu}-\mu_{\alpha}\in[-D\ldots D]$) with constant density of states $\rho_{\alpha\s}$, and taking the inverse temperatures of the leads to be equal $\beta_{\mrL}=\beta_{\mrR}=\beta$, the expression (\ref{sorttee2}) becomes
\footnotesize
\begin{equation}
\begin{aligned}
  &\begin{aligned}
  H^{(2)}_{\m{D}\m{D'}}
  \approx\frac{1}{2}\sum_{\substack{\alpha\s\\n n'\\\ket{\psi_{\D}}}}\m{P}\int_{-D}^{D} d\xi
  \Biggr[
  &t_{\alpha\s}^{n}(t_{\alpha\s}^{n'})^{*}f(\xi)
  \bra{\m{D}}\dan_{n}\ket{\psi_{\D}}\bra{\psi_{\D}}\dd_{n'}\ket{\m{D'}}
  \left(\frac{1}{\xi+\mu_{\alpha}+E^{{}}_{\m{D}}-E^{{}}_{\psi_{\D}}}+\frac{1}{\xi+\mu_{\alpha}+E^{{}}_{\m{D'}}-E^{{}}_{\psi_{\D}}}\right)\\
  +&(t_{\alpha\s}^{n})^{*}t_{\alpha\s}^{n'}\{1-f(\xi)\}
  \bra{\m{D}}\dd_{n}\ket{\psi_{\D}}\bra{\psi_{\D}}\dan_{n'}\ket{\m{D'}}
  \left(\frac{1}{-\xi-\mu_{\alpha}+E^{{}}_{\m{D}}-E^{{}}_{\psi_{\D}}}+\frac{1}{-\xi-\mu_{\alpha}+E^{{}}_{\m{D'}}-E^{{}}_{\psi_{\D}}}\right)
  \Bigg]
  \end{aligned}\\
  &\begin{aligned}
  \phantom{.....}
  \stackrel{\beta\rightarrow+\infty}{=}\frac{1}{2}\sum_{\substack{\alpha\s\\n n'\\\ket{\psi_{\D}}}}
  \Biggr[
  &t_{\alpha\s}^{n}(t_{\alpha\s}^{n'})^{*}
  \bra{\m{D}}\dan_{n}\ket{\psi_{\D}}\bra{\psi_{\D}}\dd_{n'}\ket{\m{D'}}
  \left(\ln\absB{\frac{E_{\m{D}}-E_{\psi_{\D}}+\mu_{\alpha}}{E_{\m{D}}-E_{\psi_{\D}}+\mu_{\alpha}-D}}
  +\ln\absB{\frac{E_{\m{D'}}-E_{\psi_{\D}}+\mu_{\alpha}}{E_{\m{D'}}-E_{\psi_{\D}}+\mu_{\alpha}-D}}\right)\\
  +&(t_{\alpha\s}^{n})^{*}t_{\alpha\s}^{n'}
  \bra{\m{D}}\dd_{n}\ket{\psi_{\D}}\bra{\psi_{\D}}\dan_{n'}\ket{\m{D'}}
  \left(\ln\absB{\frac{E_{\m{D}}-E_{\psi_{\D}}-\mu_{\alpha}}{E_{\m{D}}-E_{\psi_{\D}}-\mu_{\alpha}-D}}
  +\ln\absB{\frac{E_{\m{D'}}-E_{\psi_{\D}}-\mu_{\alpha}}{E_{\m{D'}}-E_{\psi_{\D}}-\mu_{\alpha}-D}}\right)
  \Bigg],
  \end{aligned}
\end{aligned}
\end{equation}
\normalsize
where $\m{P}$ denotes the principal part of the integral. Here after the second equality we have taken the zero temperature limit ($\beta\rightarrow+\infty$), which also corresponds to taking $\ket{\LRo}$ to be the zero temperature ground state of the leads. For very large bandwidth compared to other energy scales we acquire Eq.~(\ref{soehm}).

\twocolumngrid

\bibliography{bib}

\end{document}